\documentclass[twocolumn]{article}
\usepackage{nolta2012}
\usepackage{txfonts}

\usepackage{epsf,graphicx,psfrag}

\begin{document}

\title{Breathers and surface modes in oscillator chains with Hertzian interactions}

\author{Guillaume James${}^{\ast \star}$, Jes\'us Cuevas${}^\dag$ and Panayotis G. Kevrekidis${}^\ddag$}

\address{
${}^\ast$ Laboratoire Jean Kuntzmann, Universit\'e de Grenoble and CNRS\\
BP 53, 38041 Grenoble Cedex 9, France\\
${}^\star$ INRIA, Bipop Team-Project\\
ZIRST Montbonnot, 655 Avenue de l'Europe, 38334 Saint Ismier, France\\
\dag Grupo de F\'{\i}sica No Lineal,
Departamento de F\'{\i}sica Aplicada I, Escuela Polit\'{e}cnica Superior, Universidad de Sevilla\\ 
C/ Virgen de \'{A}frica, 7, 41011 Sevilla, Spain\\
\ddag Department of Mathematics and Statistics, University of Massachusetts\\ 
Amherst, Massachussets 01003-4515, USA \\
Email: guillaume.james@imag.fr, jcuevas@us.es, kevrekid@math.umass.edu
}

\maketitle

\abstract
We study localized waves in chains of oscillators 
coupled by Hertzian interactions and trapped in local potentials.
This problem is originally motivated by Newton's cradle,
a mechanical system consisting of a chain of touching beads
subject to gravity and attached to inelastic strings. 
We consider an unusual setting with local oscillations 
and collisions acting on similar time scales, a situation
corresponding e.g. to a modified Newton's cradle with
beads mounted on stiff cantilevers. Such systems
support static and traveling breathers with unusual properties, 
including double exponential spatial decay, almost vanishing Peierls-Nabarro barrier
and spontaneous direction-reversing motion. 
We prove analytically the existence of surface modes and
static breathers
for anharmonic on-site potentials and weak Hertzian interactions.
\endabstract

\vspace{1ex}

Granular media are known to display a rich dynamical behavior
originating from their complex spatial structure and different sources
of nonlinearity (Hertzian contact interactions between grains, friction, plasticity). 
In the case of granular crystals (i.e. for grains organized on a lattice), 
nonlinear contact interactions lead to different types of localized wave
phenomena 
that could be potentially used for the design of smart materials such as
acoustic diodes~\cite{chiaranick}. Among the most studied types of
excitations, solitary waves
can be easily generated by an impact at one end of a chain of touching beads
(see \cite{neste2,sb} and references therein).
In the absence of an original compression
in the chain (the so-called precompression), these solitary waves
differ from the classical KdV-type solitary waves, since they are
highly-localized (with super-exponential decay) and their
width remains unchanged with amplitude 
(see e.g. \cite{atanas}).
These properties originate from
the fully nonlinear character of the Hertzian interaction potential
$V(r)=\frac{2}{5}\, \gamma\,  ( -r )_+^{5/2}$
(with $\gamma >0$ and $(a)_+ = \rm{max}(a,0)$), which yields a
vanishing sound velocity in the absence of precompression. 

Discrete breathers
(i.e. intrinsic localized modes) form
another interesting class of excitations consisting of
time-periodic and spatially localized waveforms \cite{aubryMK,flg}.
These waves exist in diatomic granular chains
under precompression \cite{boe,theo2}
(with their frequency lying between the acoustic and optic phonon bands)
and can be generated e.g. through modulational instabilities.
However, because precompression suppresses the fully nonlinear character
of Hertzian interactions, these excitations inherit the usual properties of
discrete breathers, i.e. their spatial decay is exponential
and their width diverges at vanishing amplitude
(for frequencies close to the bottom of the optic band). 

The situation is sensibly different for granular systems without precompression.
In that case, localized oscillations can be generated on short transients 
in the form of transitory defect modes induced by a mass impurity 
(see \cite{s12} and references therein)
but never occur {\em time-periodically} as proved in \cite{jkc}. 
Indeed, uncompressed granular chains are described by the Fermi-Pasta-Ulam lattice
with Hertzian interactions 
\begin{equation}
\label{eqm1}
\ddot{y}_n  = V^\prime (y_{n+1} -y_n) - V^\prime (y_{n} -y_{n-1})
\end{equation}
(or spatially inhomogeneous variants thereof), where
$y_n(t)$ denotes the $n$th bead
displacement from its reference position. 
For all $T$-periodic solutions of (\ref{eqm1}), 
the average interaction forces
$\int_{0}^{T}{V^\prime (y_{n+1} -y_n)\, dt}$ are independent of $n$
(this is immediate by integrating (\ref{eqm1})). Consequently,
localized oscillations
would yield a vanishing average interaction force between grains, 
which is impossible since Hertzian interactions are
repulsive under contact and vanish otherwise.

\vspace{1ex}

\begin{figure}[h]
\begin{center}
\includegraphics[scale=0.1]{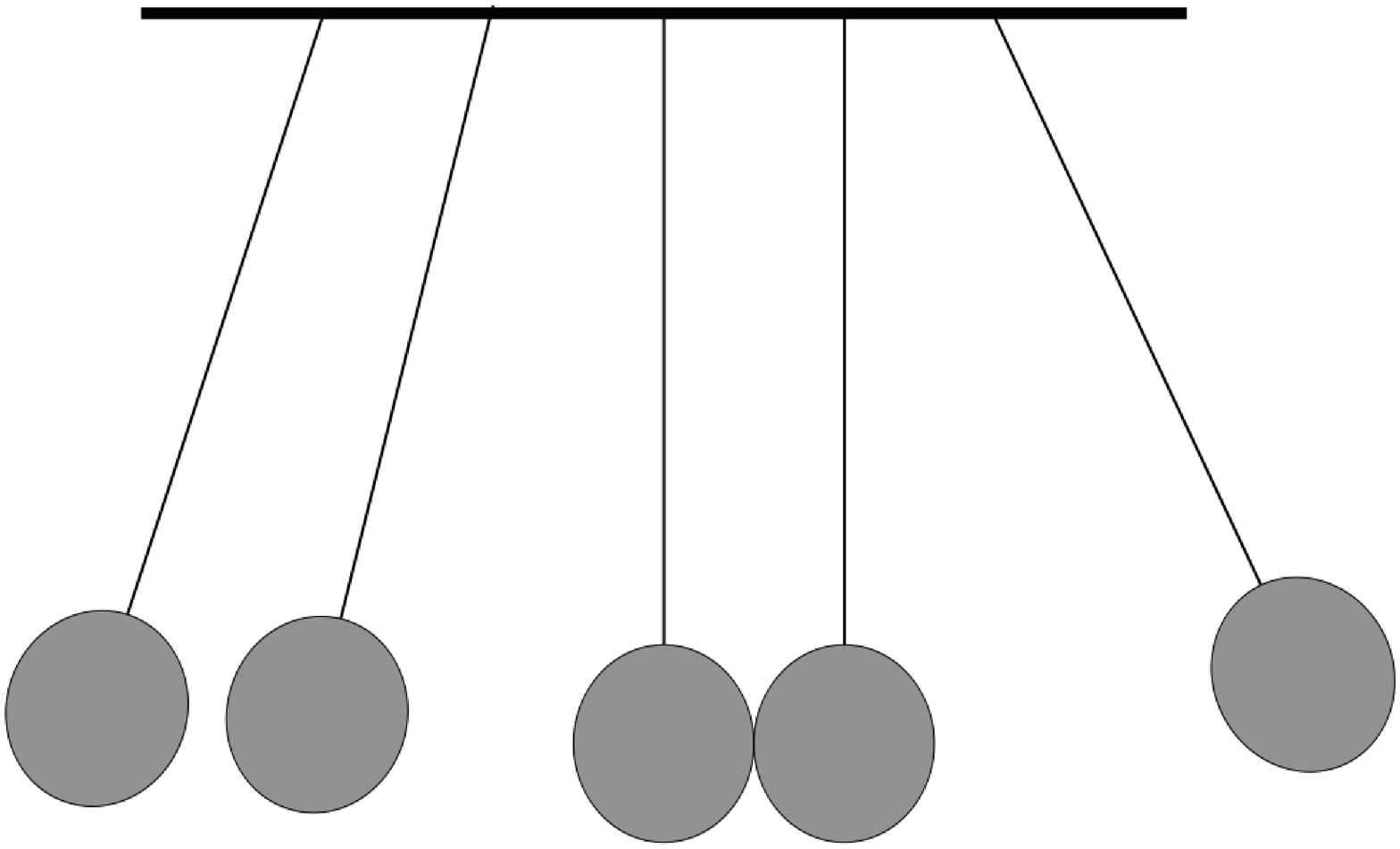}
\includegraphics[scale=0.1]{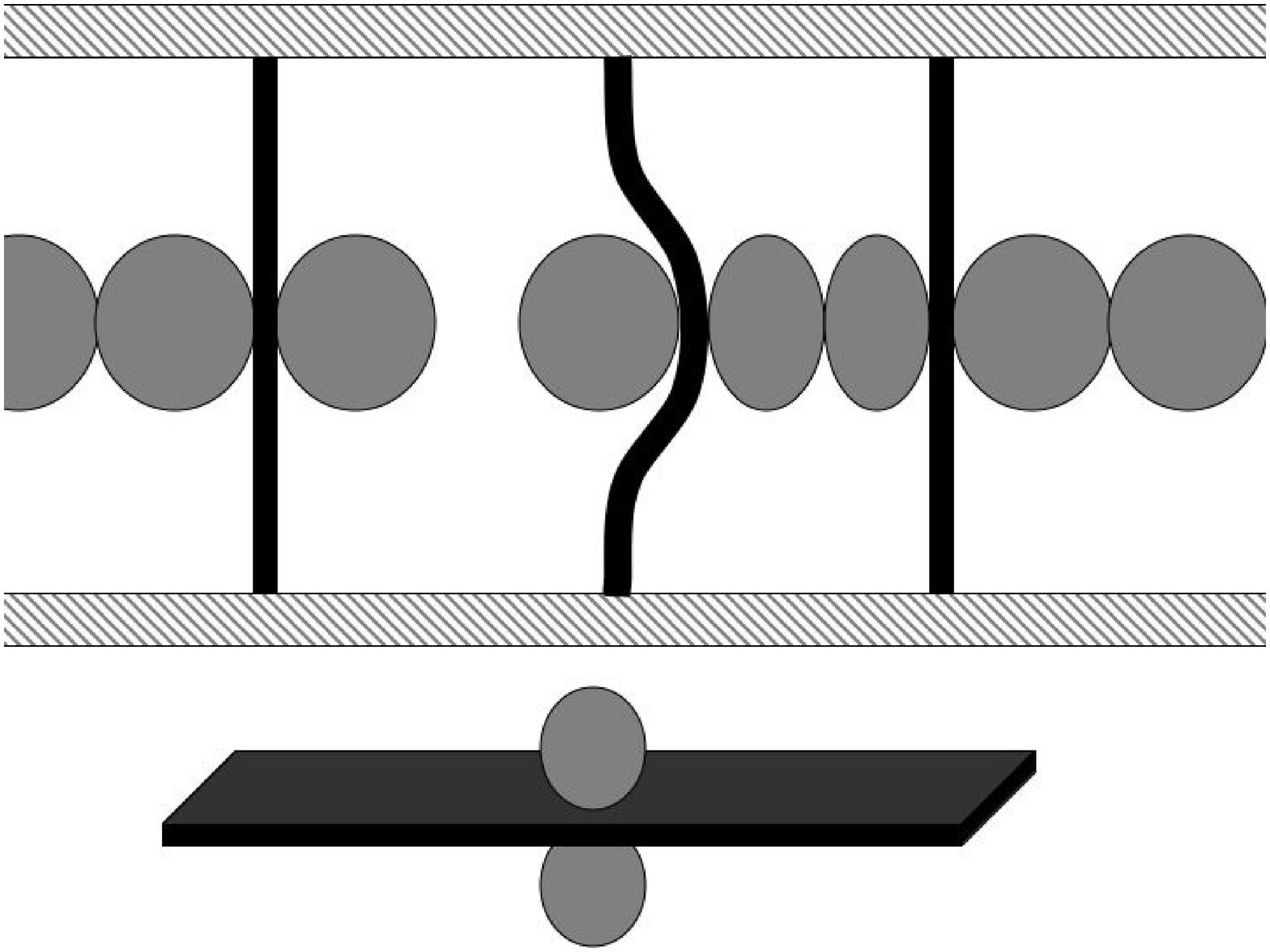}
\end{center}
\caption{\label{boules}
Left~: prototypical Newton's cradle
consisting of a chain of beads attached to pendula.
Right~:
array of clamped cantilevers decorated by spherical beads.}
\end{figure}

In contrast to the above picture, we have numerically established in \cite{jkc} 
the existence of time-periodic localized oscillations in Hertzian chains with 
symmetric local potentials described by the system
\begin{equation}
\label{eqm}
\ddot{y}_n  + W^\prime (y_n) = V^\prime (y_{n+1} -y_n) - V^\prime (y_{n} -y_{n-1})
\end{equation}
where $W(y)=\frac{1}{2}\, y^2+\frac{s}{4}\,  y^4$
and $s\in \mathbb{R}$ measures the local anharmonicity.
System (\ref{eqm})
describes small amplitude waves in
a Newton's cradle \cite{jamesc} (figure \ref{boules}, left)
or other mechanical systems
consisting of beads mounted e.g. on an elastic matrix \cite{sh12} or cantilevers \cite{jkc} (figure \ref{boules}, right). 
In the last two cases, local oscillations and collisions between beads can occur on similar time scales
for realistic material parameter values, which allows for breathing dynamics to take place.
Static and moving breathers can be generated from standard initial conditions
such as a localized impact \cite{jkc,sh12} or perturbations of unstable
periodic traveling waves \cite{jamesc}. 

\begin{figure}[h]
\psfrag{5}[0.9]{\tiny 5}
\psfrag{10}[0.9]{\tiny 10}
\psfrag{15}[0.9]{\tiny 15}
\psfrag{20}[0.9]{\tiny 20}
\psfrag{25}[0.9]{\tiny 25}
\psfrag{30}[0.9]{\tiny 30}
\psfrag{1.0e-04}[0.9]{\tiny $10^{-4}$}
\psfrag{0.0e+00}[0.9]{\tiny ~~~~~~~~~0}
\psfrag{-1.0e-04}[0.9]{\tiny $-10^{-4}$}
\psfrag{5.0e-05}[0.9]{\tiny ~}
\psfrag{-5.0e-05}[0.9]{\tiny ~}
\psfrag{Initial positions : numerical solution (marks), continuum approximation (full line)}[0.9]{\tiny ~}
\psfrag{8e-05}[0.9]{\tiny $\hspace*{-3ex}8\cdot 10^{-5}$}
\psfrag{-8e-05}[0.9]{\tiny $\hspace*{-3ex}-8\cdot 10^{-5}$}
\psfrag{0e+00}[0.9]{\tiny ~~~~~~~0}
\psfrag{6e-05}[0.9]{\tiny ~}
\psfrag{-6e-05}[0.9]{\tiny ~}
\psfrag{4e-05}[0.9]{\tiny ~}
\psfrag{-4e-05}[0.9]{\tiny ~}
\psfrag{2e-05}[0.9]{\tiny ~}
\psfrag{-2e-05}[0.9]{\tiny ~}
\psfrag{n}[0.9]{ $n$}
\psfrag{x_n(0)}[1][Bl]{ $y_n (0)$}
\begin{center}
\includegraphics[scale=0.204]{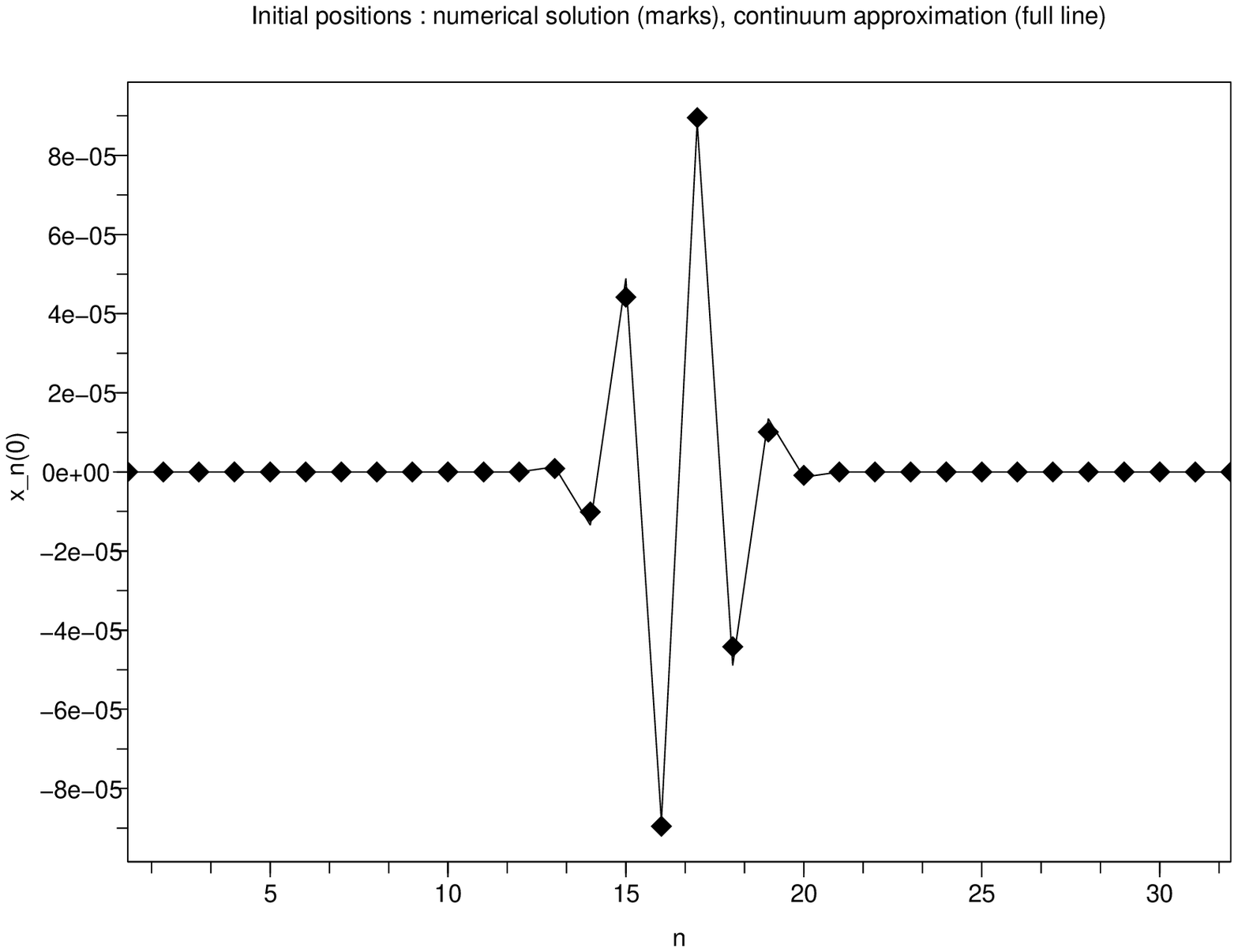}
\includegraphics[scale=0.204]{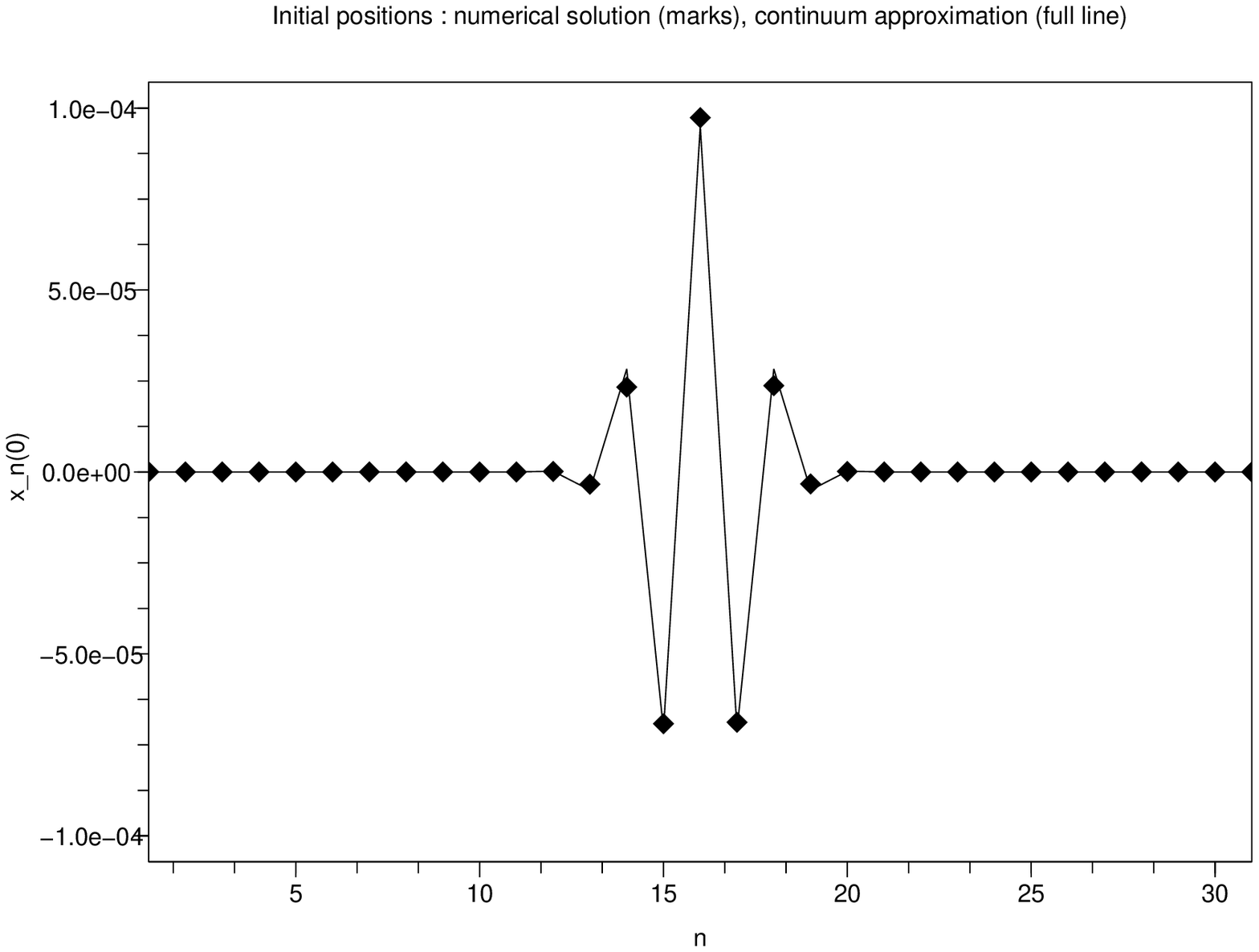}
\end{center}
\caption{\label{breathers}
Bond-centered (left plot) and site-centered (right plot)
breather solutions of (\ref{eqm}) for $\omega_b=1.01$
(potential parameters are $\gamma =1$ and $s=0$).
The numerical solutions 
(marks) are compared to the
quasi-continuum approximations $y_n^{(1)}$, $y_n^{(2)}$ (continuous lines).}
\end{figure}

Let us summarize the results of \cite{jkc}.
Two families of static breather solutions of (\ref{eqm}) 
(parametrized by their frequency $\omega_b >1$)
have been computed using the Gauss-Newton method and path-following.
The first one consists of bond-centered breathers
satisfying $y_n(t)=-y_{-n+1}(t):=\mathcal{S}_1y_n(t)$, and the second corresponds to site-centered breathers with
$y_n(t)= -y_{-n}(t+T_b/2):=\mathcal{S}_2y_n(t)$, where $T_b$ denotes the breather period
(each solution family is invariant by a symmetry $\mathcal{S}_i$ of 
(\ref{eqm})).
Profiles of both breather types 
are given in figure \ref{breathers} for a particular
frequency close to unity (i.e. the linear frequency of local oscillators), a limit where the
breather amplitude vanishes. Near this limit, 
quasi-continuum approximations of the breather profiles can
be derived \cite{jkc}, namely
\begin{equation}
\label{ansatzapprox1}
y_n^{(1)} (t)= 2 \epsilon\, 
(-1)^{n}\, [g(n)+g(n-1)] 
\, \cos{(\omega_{\rm{b}} t )}
\end{equation}
for bond-centered breathers and
\begin{equation}
\label{ansatzapprox2}
y_n^{(2)} (t)= 2 \epsilon\, 
(-1)^{n}\, [g(n+\frac{1}{2})+g(n-\frac{1}{2})]
\, \cos{(\omega_{\rm{b}} t )}
\end{equation}
for site-centered ones, 
where $\omega_{\rm{b}} =1+ \frac{\epsilon^{1/2}}{2\tau_0}$ ($\tau_0 \approx 1.545$),
$g(x)=\big(\frac{3}{10}\big)^3\, \cos^6{\big(  \frac{x}{3} \big)} \mbox{ for } |x|\leq \frac{3\pi}{2}$ and
$g=0$ elsewhere. Figure \ref{breathers} compares the numerical solutions with the above
approximations for $\omega_b \approx 1$, showing excellent agreement. Obviously
discrepancies appear at larger amplitude, since 
approximations (\ref{ansatzapprox1}) and (\ref{ansatzapprox2}) retain only one Fourier mode
and are independent of the local anharmonicity 
(which is dominated by the Hertzian nonlinearity at small amplitude).
The above approximations possess compact supports
(with a width independent of $\omega_b$), but
the breathers computed numerically do not share this property strictly speaking and
instead  display super-exponential localization (this is
in analogy with the case of homogeneous polynomial 
interaction potentials, see section 4.1.3 of \cite{flg}).

\begin{figure}[h]
\begin{center}
\includegraphics[scale=0.218]{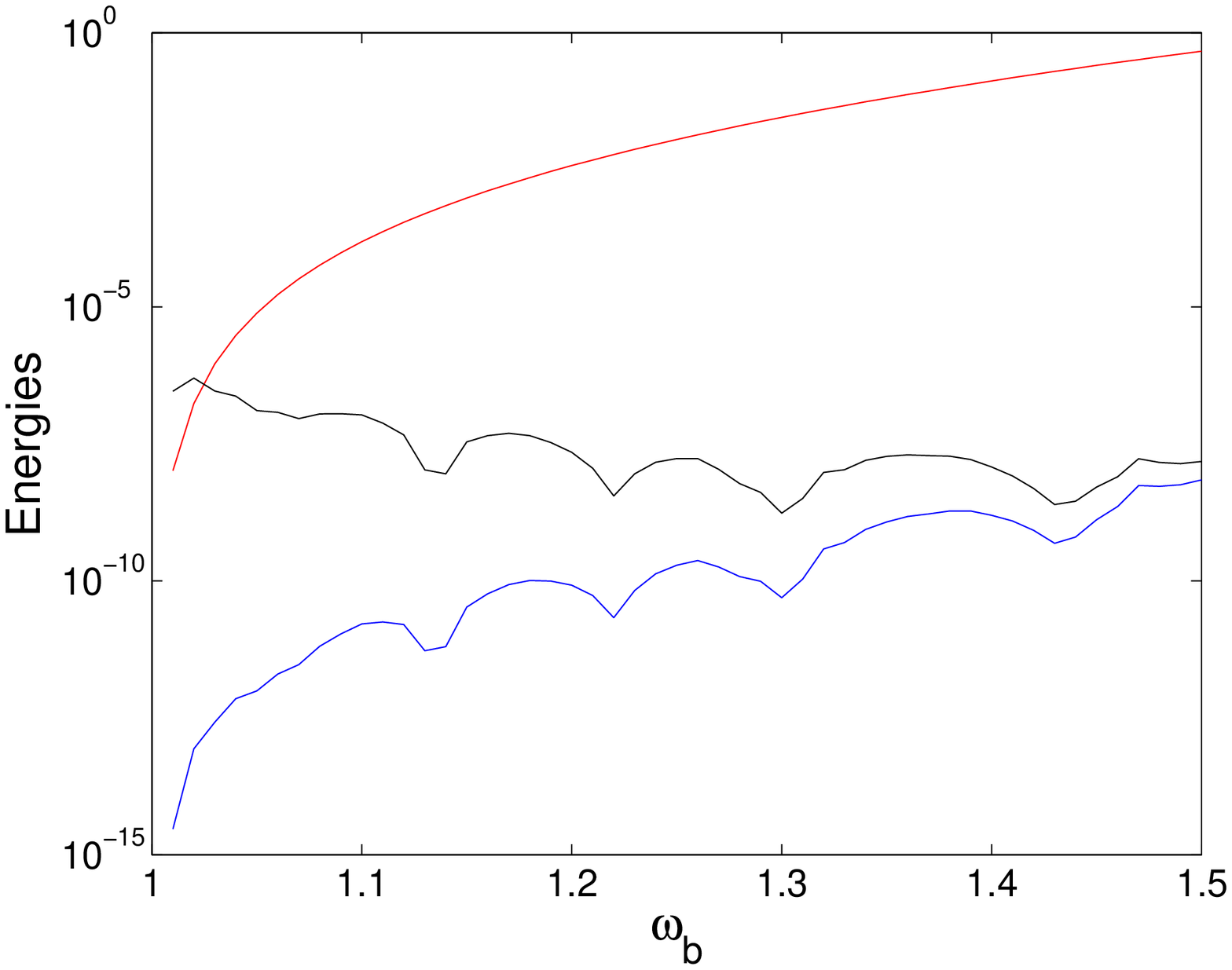}
\includegraphics[scale=0.218]{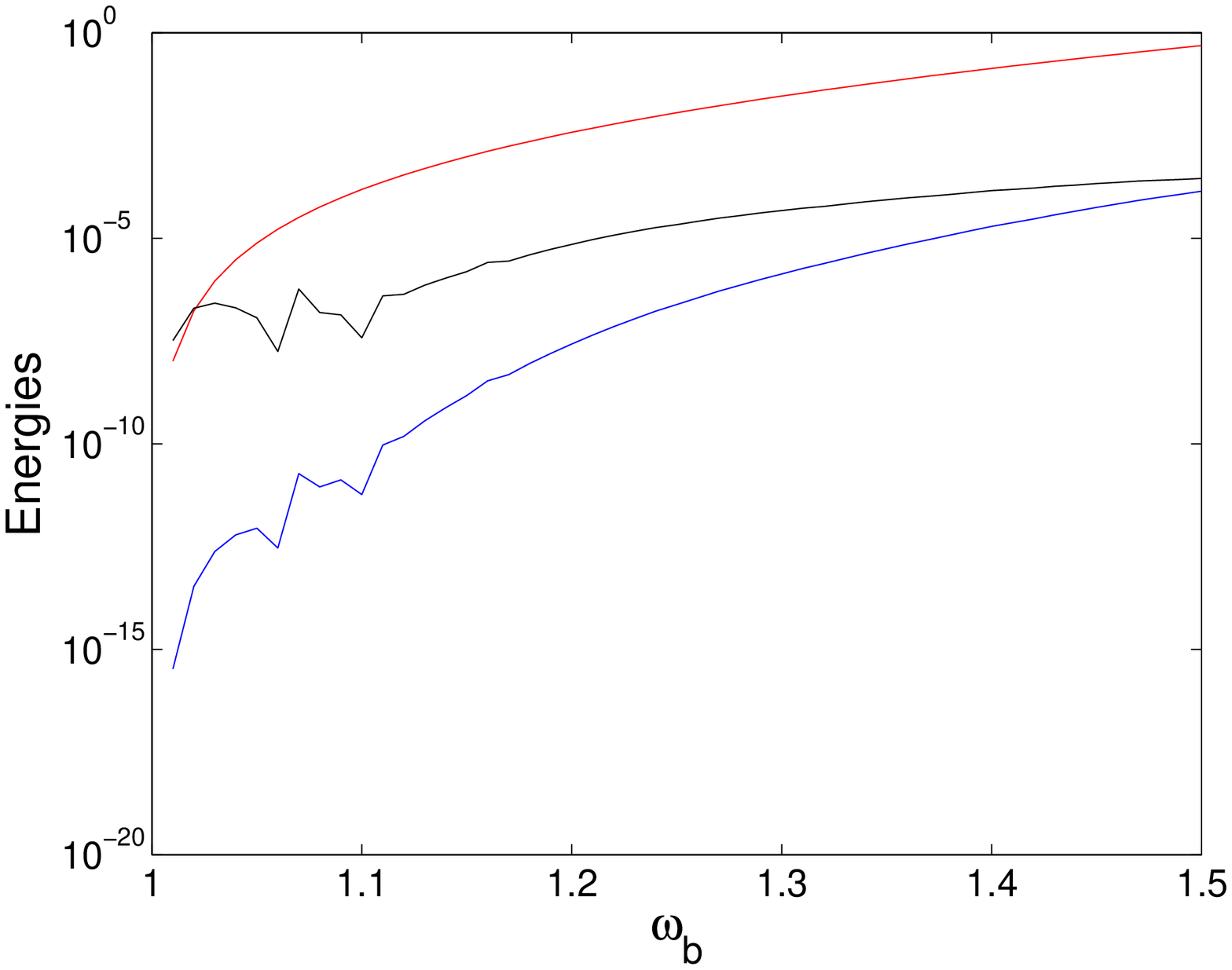}
\end{center}
\caption{\label{pnb}
Breather energies versus frequency $\omega_b$,
when the on-site potential is harmonic ($s=0$, left plot) or
anharmonic ($s=-1/6$, right plot) and $\gamma =1$.
The red curves give the energy ${\mathcal H}_{bc}$ of bond-centered breathers,
the blue curves the approximate Peierls-Nabarro barrier $E_{PN}$
(energy difference bewteen site- and bond-centered breathers)
and the black curves the relative energy ratio $E_{PN}/{\mathcal H}_{bc}$.
For small amplitude breathers (i.e. $\omega_b \approx 1$),
the different values of $s$ yield comparable values of ${E}_{PN}$. Clearly
${E}_{PN}$ increases with
breather amplitude but remains very small in this parameter range
(e.g. ${E}_{PN}$ is close to $10^{-4}$ for $\omega_b =1.5$ and $s=-1/6$).
The harmonic case yields even smaller barriers,
by $3-4$ orders of magnitude for $\omega_b=1.3$.}
\end{figure}

\begin{figure}[h]
\psfrag{t}[0.9]{$t$}
\psfrag{n}[0.9]{\hspace*{5.2ex}$n$}
\begin{center}
\includegraphics[scale=0.219]{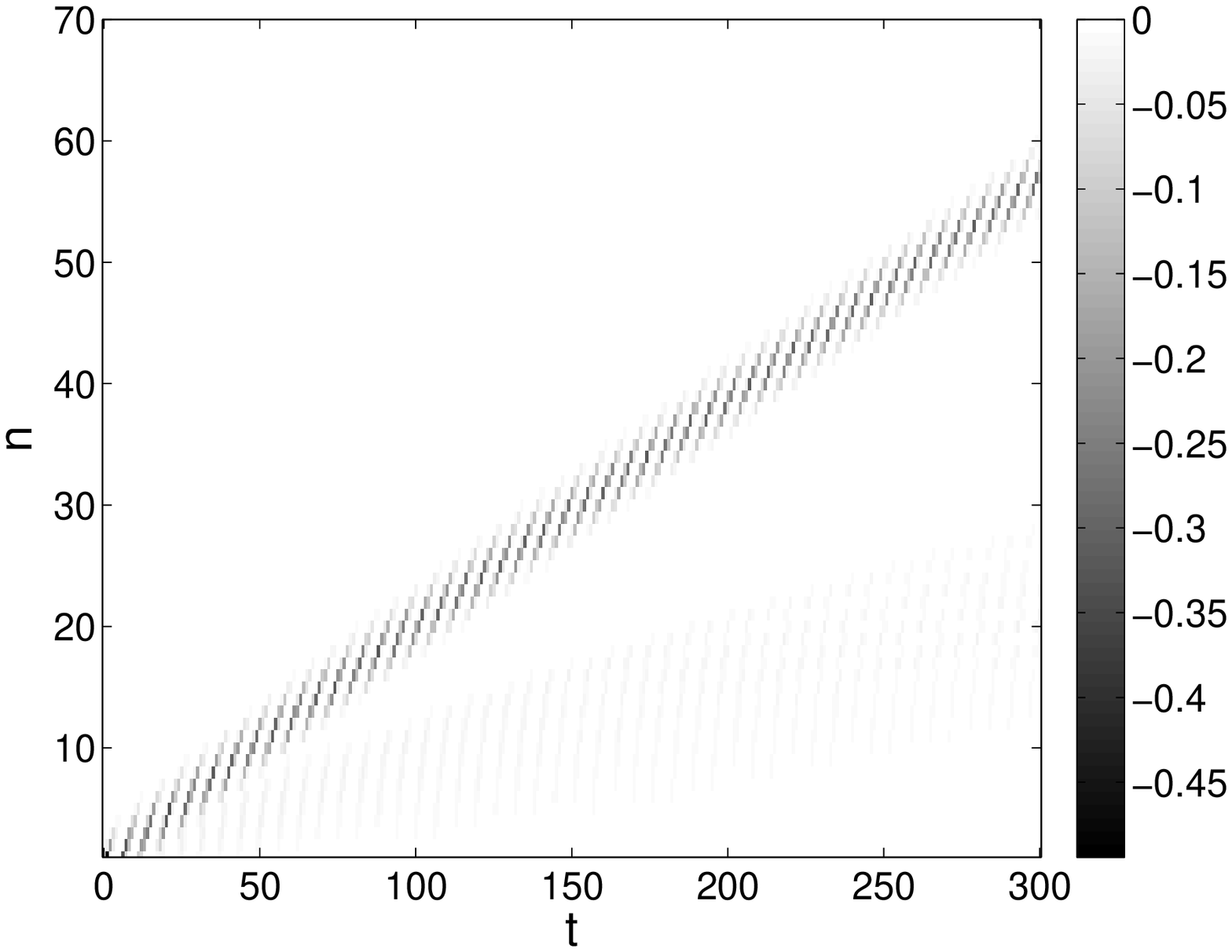}
\includegraphics[scale=0.219]{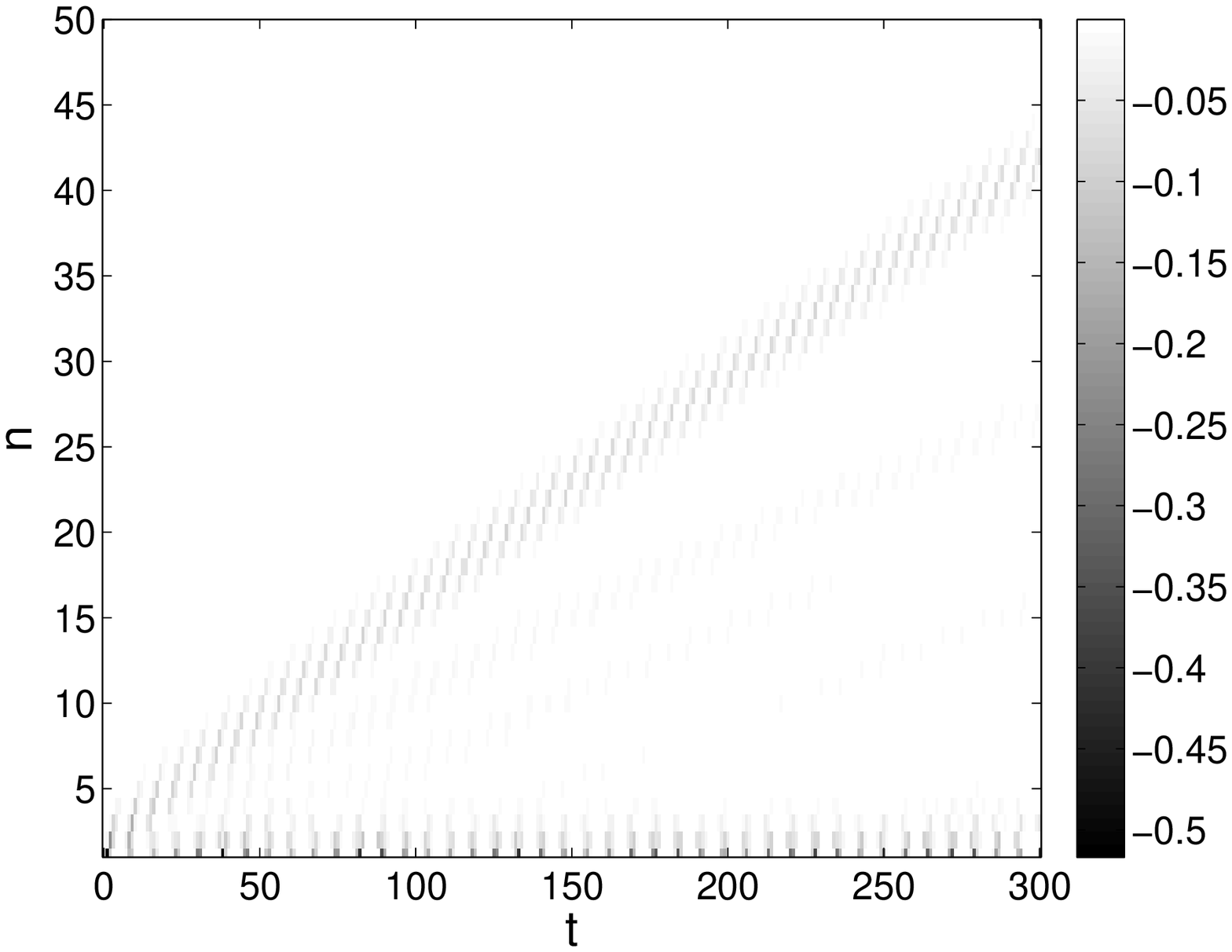}
\end{center}
\caption{\label{surfmode}
Space-time diagrams showing the interaction forces
$f_n = -(y_n - y_{n+1})_+^{3/2}$ in system (\ref{eqm})
with $\gamma =1$ and free end boundary conditions.
Forces are represented
in grey levels, white corresponding to vanishing interactions (beads not in contact)
and black to a minimal negative value of the contact force.
The initial condition is $y_n(0)=0$ for $n\geq 1$,
$\dot{y}_n(0)= 0$ for $n\geq 2$, $\dot{y}_1(0)= 0.94$. 
For an harmonic local potential ($s=0$, left plot), a traveling breather is generated. 
For a soft local potential ($s=-0.7$, right plot), one observes in addition
the excitation of a surface mode.
}
\end{figure}

Dynamical simulations of \cite{jkc} indicate that extremely small perturbations of the static
breathers can lead to their translational motion
(generating a so-called traveling breather),
for anharmonic on-site potentials,
but more critically even in the harmonic case. 
This phenomenon is linked with an extremely small difference between the energies ${\mathcal H}$ of 
site- and bond-centered breathers having the same frequency
(the so-called approximate Peierls-Nabarro barrier).
Typical values of breather energies are given in figure \ref{pnb},
where ${\mathcal H}=\sum_{n}{e_n}$
and $e_n=\frac{1}{2}\, \dot{y}_n^2+W( y_n ) + V(y_{n+1}-y_n)$.
Another manifestation of the high breather mobility is the systematic formation
of a travelling breather after an impact at one end of the oscillator chain
(see figure \ref{surfmode}).

\begin{figure}[h]
\psfrag{0}[0.9]{\tiny ~0}
\psfrag{10}[0.9]{\tiny ~}
\psfrag{20}[0.9]{\tiny 20}
\psfrag{30}[0.9]{\tiny ~}
\psfrag{40}[0.9]{\tiny 40}
\psfrag{50}[0.9]{\tiny ~}
\psfrag{60}[0.9]{\tiny 60}
\psfrag{70}[0.9]{\tiny ~}
\psfrag{80}[0.9]{\tiny 80}
\psfrag{90}[0.9]{\tiny ~}
\psfrag{100}[0.9]{\tiny 100}
\psfrag{0.80}[0.9]{\tiny 0.8}
\psfrag{0.00}[0.9]{\tiny ~~~~~0}
\psfrag{-1.0}[0.9]{\tiny -1}
\psfrag{0.60}[0.9]{\tiny ~}
\psfrag{0.40}[0.9]{\tiny ~}
\psfrag{0.20}[0.9]{\tiny ~}
\psfrag{-0.8}[0.9]{\tiny ~}
\psfrag{-0.6}[0.9]{\tiny ~}
\psfrag{-0.4}[0.9]{\tiny ~}
\psfrag{-0.2}[0.9]{\tiny ~}
\psfrag{t}[0.9]{\hspace*{-1.5ex}$t$}
\psfrag{t=241.52}[0.9]{\tiny ~}
\psfrag{bead velocities}[0.9]{\tiny ~}
\psfrag{n}[0.9]{ $n$}
\psfrag{x}[1][Bl]{~~~$\dot{y}_n(t) $}
\begin{center}
\includegraphics[scale=0.201]{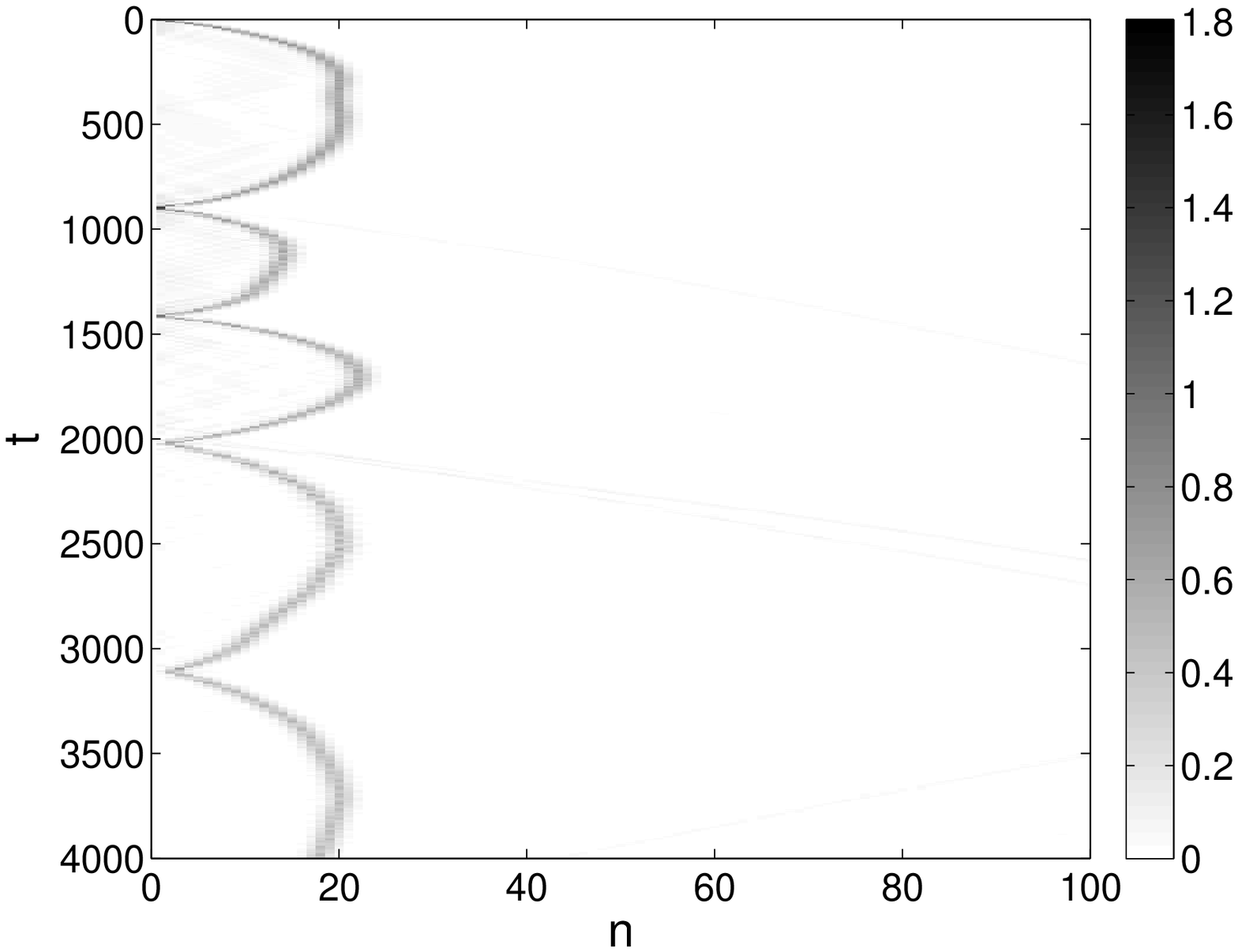}
\includegraphics[scale=0.221]{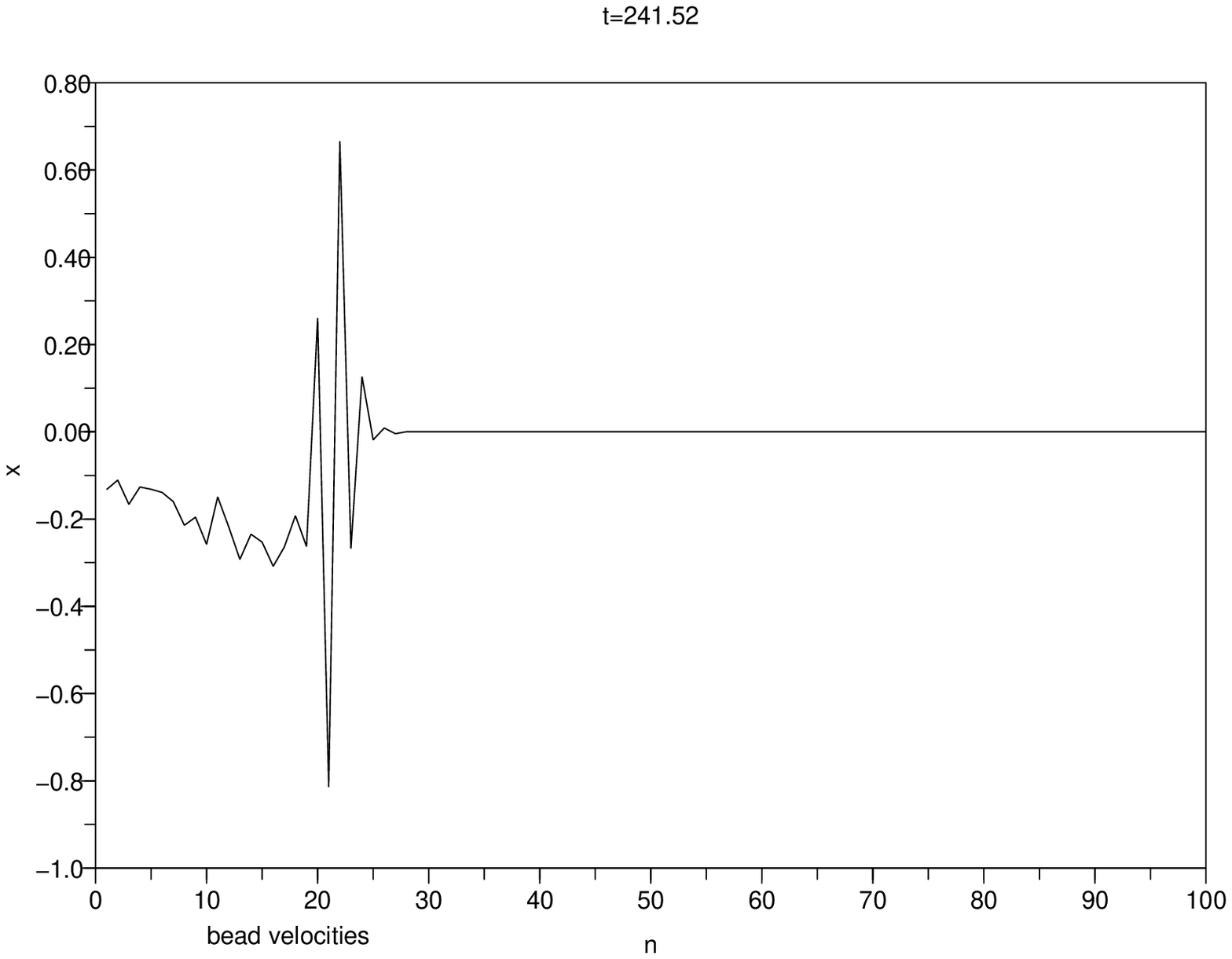}
\end{center}
\caption{\label{impactspos}
Left~: space-time diagram of the energy density $e_n$ for
system (\ref{eqm}) with $\gamma =1$ and a hard anharmonic local potential ($s=1$).
The initial condition is the same as in figure \ref{surfmode} except $\dot{y}_1(0)=1.9$.
Right~: corresponding particle velocities close to direction-reversing 
($t\approx 241$).}
\end{figure}

Different phenomena have been identified in \cite{jkc}
depending on the softening or hardening character of the local potential $W$.
Firstly, the stability of both site-centered and bond-centered
breathers is critically dependent on the strength (and sign) of the anharmonicity
(we refer to \cite{jkc} for more details). Secondly, 
depending whether $s<0$ or $s>0$,
the occurence of
surface mode excitations (i.e. oscillations localized near a boundary)
or direction-reversing traveling breathers was observed after an impact. 
Both phenomena are illustrated by figures
\ref{surfmode} and \ref{impactspos}.
The origin of direction reversal is still unclear at the present stage,
although we think it might originate from the interaction between
the traveling breather and nonlinear waves confined between the breather
and the boundary (figure \ref{impactspos}, right plot).

\begin{figure}[h]
\begin{center}
\includegraphics[scale=0.217]{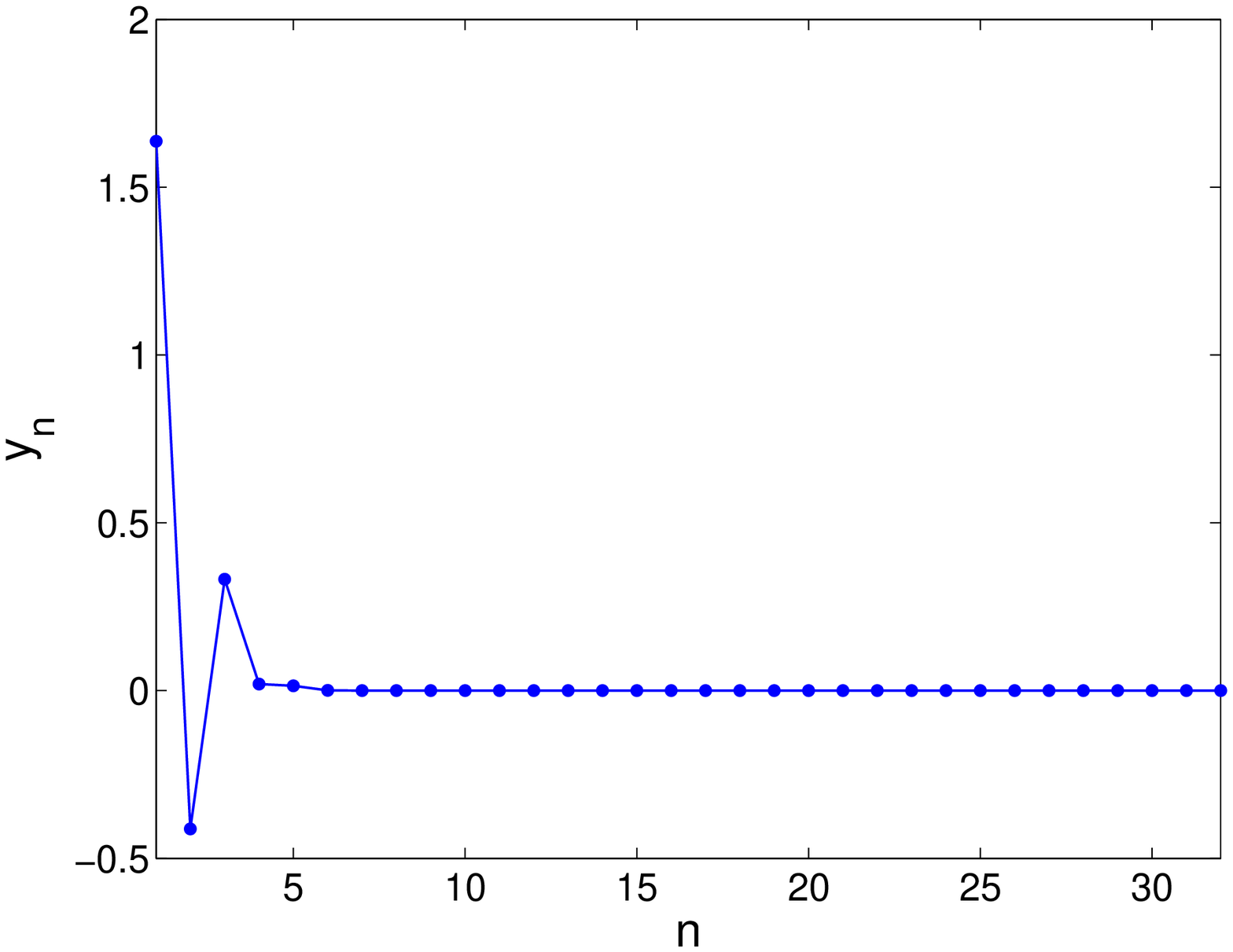}
\includegraphics[scale=0.217]{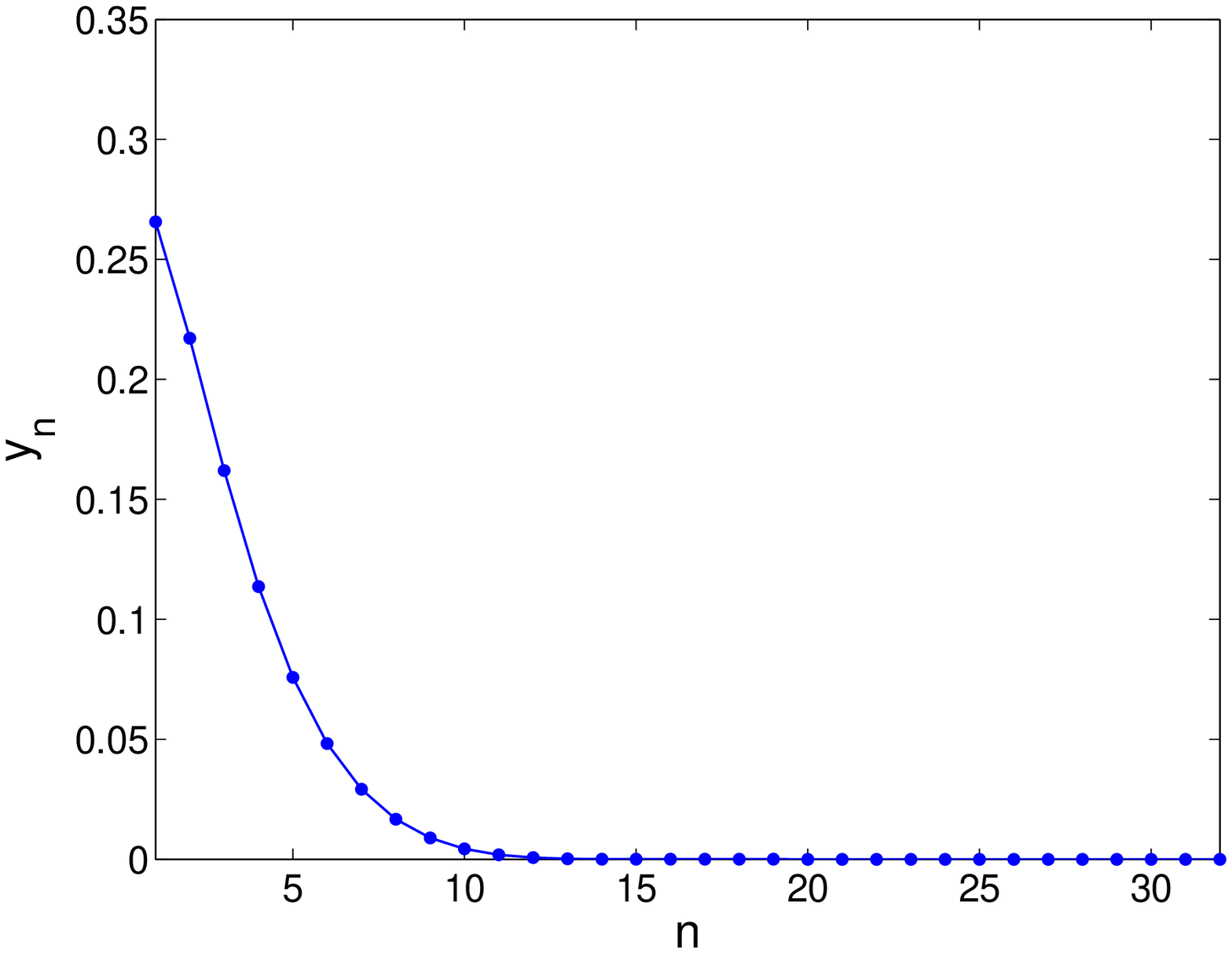}
\includegraphics[scale=0.225]{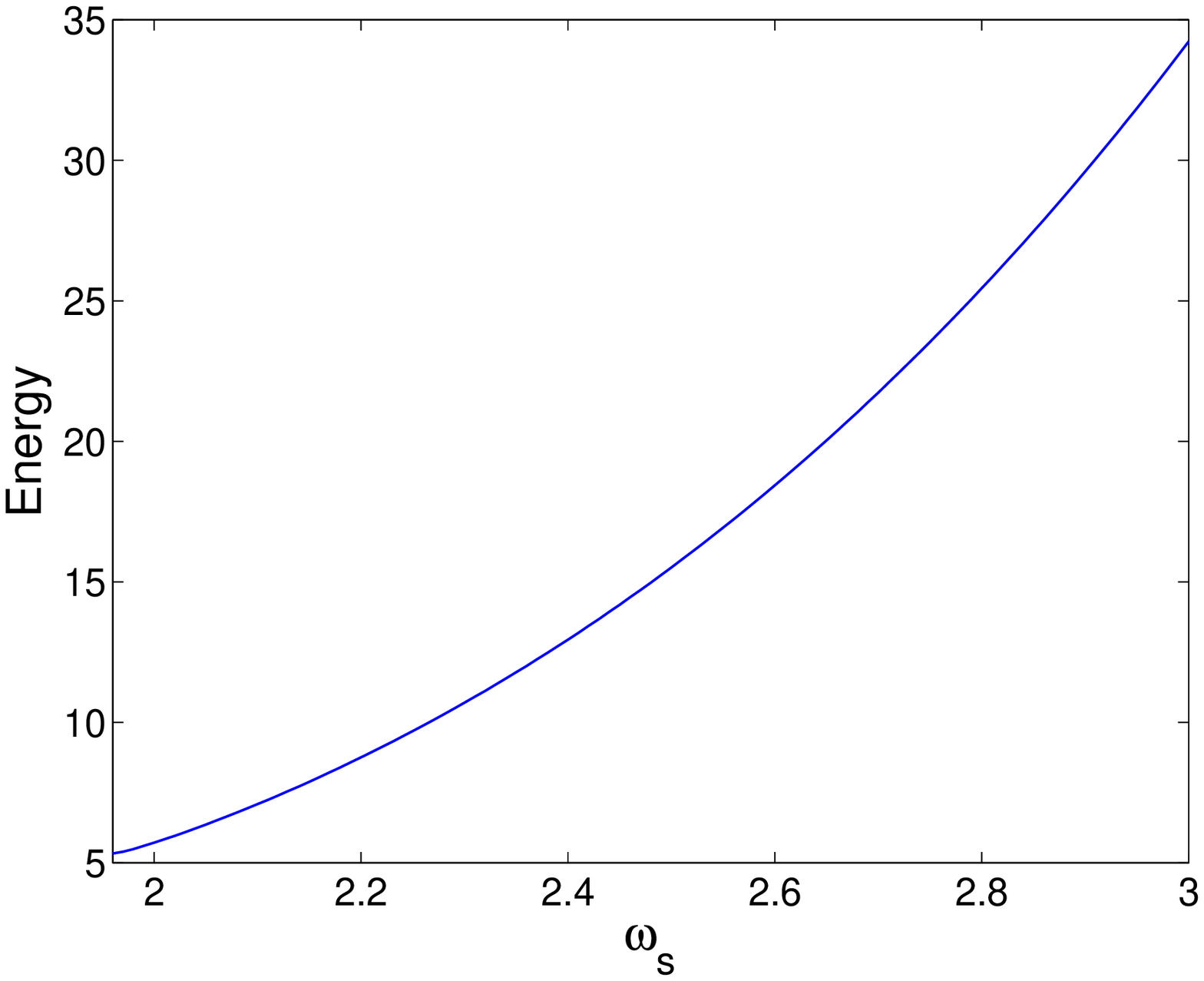}
\includegraphics[scale=0.225]{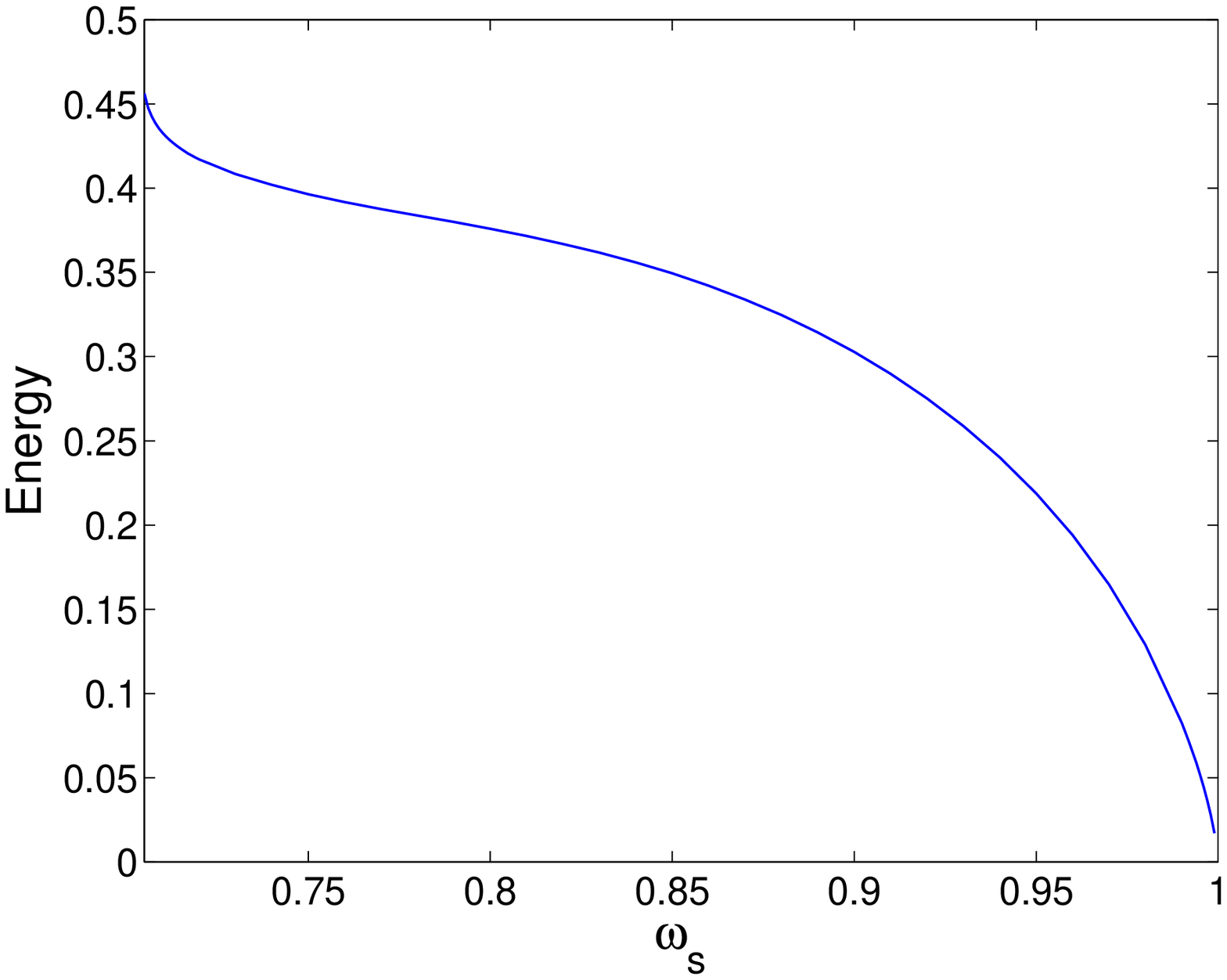}
\end{center}
\caption{\label{surfhardsoft}
Top : profile of a surface mode at the instant of maximal amplitude,
for a hard local potential ($s=1$, $\omega_s=2$, left plot)
and a soft one ($s=-0.7$, $\omega_s=0.99$, right plot).
Bottom~: energy of the surface mode versus frequency $\omega_s$,
for $s=1$ (left plot) and $s=-0.7$ (right plot).}
\end{figure}

In order to understand the absence of surface mode excitation observed in figure \ref{impactspos}
for $s=1$, we now analyze the properties of surface modes computed by the Gauss-Newton method
(all numerical computations are performed for $\gamma =1$). The results are shown in figure 
\ref{surfhardsoft} (left column). For $s=1$, surface modes exist for frequencies $\omega_s$ lying above a critical 
frequency $\omega_{\rm min} \approx 1.96$, where a pair of Floquet eigenvalues converges towards unity
(spectra not shown here). For $\omega_s \approx \omega_{\rm min}$ these solutions are spectrally stable
and one can observe an energy threshold. This explains why a surface mode is
not observed in figure \ref{impactspos} where the energy of the initial excitation is well below the excitation threshold.  
This situation contrasts with the case of soft local potentials illustrated in figure \ref{surfhardsoft} (right column).
For $s=-0.7$, surface modes exist for $\omega_s\in [0.705,1)$. 
They are spectrally stable for $\omega_s \approx 1$ but oscillatory instabilities appear at smaller
frequencies (spectra not shown here). When $\omega_s \rightarrow 1$, their energy
and amplitude vanish, opening the possibility of exciting
such modes for arbitrarily small initial velocities of the first particle.

Having numerically analyzed the properties of surface modes of (\ref{eqm}) 
when $\gamma$ 
(the stiffness constant of Hertzian interactions) equals unity,
we now prove the existence of time-periodic and spatially localized solutions
when $\gamma$ is small. The small coupling (anticontinuum) limit was introduced in
reference \cite{aubryMK} (and considerably generalized in \cite{sepmac}) and applies
both to breathers and surface modes.
From a physical point of view, this parameter regime can be realized
with an array of clamped cantilevers (see figure \ref{boules})
decorated by spherical beads made from a sufficiently soft material (e.g. rubber).
In addition, the anticontinuum limit requires an anharmonic local potential $W$, i.e. $s\neq 0$.

We consider system (\ref{eqm}) for
$n\in \Gamma$, with $\Gamma = \mathbb{Z}$ for discrete breathers (case of a doubly infinite chain)
and $\Gamma =  \mathbb{N}_0$ for surface modes
(semi-infinite chain, with free-end boundary conditions at $n=0$).
For $\gamma =0$, system (\ref{eqm}) admits localized periodic solutions satisfying
$y_n = 0\ \forall n\neq 0$, 
$\ddot{y}_0  + W^\prime (y_0) =0$ and
$y_0 (t+T_0)=y_0(t)$.
Setting $\dot{y}_0(0)=0$, 
$a:=y_0(0)>0$ and $T_0\equiv T_0 (a)$, it is a classical result that $s\, T_0^\prime (a) <0$. Consequently,
the time periodic oscillations $y_0$ can be parametrized by their period $T_0$, with
$T_0<2\pi$ for $s>0$ and $T_0>2\pi$ for $s<0$. 
We shall denote by $Y_{0,T_0}:=(y_n)_n$ the even $T_0$-periodic solution constructed above for $\gamma =0$.
Now let us assume $T_0 \notin 2\pi \mathbb{N}$
(this condition is automatically satisfied for $s>0$). Thanks to
this nonresonance condition, one can apply the persistence theorem for
normally non-degenerate discrete breathers proved in reference \cite{sepmac} (section 3.5),
which readily applies to system (\ref{eqm}) since $V \in C^2 (\mathbb{R})$ has sufficient smoothness. 
More precisely, for $\gamma$ small enough in system (\ref{eqm}), this result ensures the existence of
a family of localized solutions $Y_{\gamma,T}(.+\phi )$ parametrized by their period $T \approx T_0$
(since $T_0^\prime \neq 0$) and an arbitrary phase $\phi$. In addition the perturbed and unperturbed solutions
are close when $\gamma$ is small, i.e. one has $Y_{\gamma,T}=Y_{0,T}+O(\gamma )$
in $C^2(S^1;\ell_2(\Gamma))$ (we denote $S^1 = \mathbb{R}/(T\mathbb{Z})$).
Note that the corresponding breather solutions are site-centered,
due to the local uniqueness of $\gamma \mapsto Y_{\gamma,T}$ and the
invariance of (\ref{eqm}) by the symmetry $\mathcal{S}_2$.
Similarly, starting from a bond-centered breather for $\gamma=0$ (with $y_1 = -y_0$, $y_n=0$ elsewhere)
allows one to obtain bond-centered breathers for $\gamma \approx 0$. More generally, multibreathers
having many excited sites can be constructed in the same way at small coupling.

Now let us estimate the spatial decay of breathers and surface modes.
Consider a $T$-periodic solution of (\ref{eqm}) satisfying the nonresonance condition $T \notin 2\pi \mathbb{N}$.
We assume this solution is spatially localized, i.e. its supremum norm 
$v_n = \| y_n\|_\infty:=\sup_{t\in [0,T]}{|y_n (t)|}$ satisfies
$\lim_{|n|\rightarrow +\infty}{v_n}=0$. In order to estimate $v_n$, we rewrite (\ref{eqm}) in the form
\begin{equation}
\label{eqmbis}
y_n = L^{-1}\, \left( \, \gamma\,  ( y_{n-1} -y_n)_+^{3/2} -\gamma\,  ( y_{n} -y_{n+1})_+^{3/2}-s\, y_n^3 \, \right),
\end{equation}
where $L=\partial^2_t +I\, : \, C^2(S^1;\mathbb{R}) \rightarrow C^0(S^1;\mathbb{R})$ is invertible thanks to the nonresonance
assumption.
By using (\ref{eqmbis}),
there exists $C>0$ and $m \in \mathbb{N}_0$ such that for $|n| \geq m+1$ we have
$v_n \leq \frac{C}{3}\, (v_{n+1}^{3/2}+v_{n}^{3/2}+v_{n-1}^{3/2})$.
Introducing the sequence $\epsilon_j = \sup_{|k|\geq j}{v_k}$, it follows
that $\epsilon_j \leq {C}\epsilon_{j-1}^{3/2}$ for $j \geq m+1$, hence
one obtains by induction 
\begin{equation}
\label{decay}
\epsilon_j \leq C^{-2}\, \lambda^{(3/2)^{j}}, \ \ \ \lambda = \left( C^2\, \epsilon_m \right)^{(2/3)^{m}}, \ \ \
j \geq m.
\end{equation}
Since $\lim_{j\rightarrow +\infty}{\epsilon_j}=0$, we can fix $m$ such that $\lambda <1$.
Estimate (\ref{decay})
implies $\| y_n\|_\infty \leq  \epsilon_{|n|}\leq C^{-2}\, \lambda^{(3/2)^{|n|}}$ for $|n| \geq m$,
i.e. the solution decays doubly exponentially at infinity.

\vspace{1ex}

As a conclusion, we have reviewed recent results of \cite{jkc}
on the numerical observation of breathers and surface modes
in granular chains with stiff local potentials. 
We have found the existence of an energy threshold for hard surface modes,
which explains their non-excitation by moderate impacts reported in \cite{jkc}.
We have obtained doubly exponential decay estimates for
breathers and surface modes and proved their existence
for anharmonic local potentials when $\gamma \approx 0$.
The persistence of these solutions for larger values of $\gamma$
is numerically observed (for $\gamma=1$) but
remains an open question from an analytical point of view.
In addition, the spontaneous direction-reversing motion
of discrete breathers occuring for hard local potentials and
the smallness of their Peierls-Nabarro energy barrier
has yet to be understood.

\section*{Acknowledgments}

\vspace*{-1.5ex}

{\small J.C. acknowledges support from the MICINN project FIS2008-04848 \&
P.G.K. from the NSF
(grant CMMI-1000337), from the A.S. Onassis 
Public Benefit Foundation (grant RZG 003/2010-2011)
and from AFOSR (grant FA9550-12-1-0332).}

\end{document}